\documentclass[prl,twocolumn,showpacs]{revtex4-1}

\usepackage{amssymb,amsmath}

\usepackage{graphicx}
\usepackage{color}
\usepackage{graphicx}
\newcommand{\vicente}[1]{{ #1}}

\begin{document}
\title{Instabilities in granular binary mixtures \vicente{at moderate densities}}
\author{Peter P. Mitrano,$^{1}$ Vicente Garz\'o,$^{2}$ and Christine M. Hrenya $^{1}$}
\affiliation{$^{1}$Department of Chemical and Biological Engineering, University of Colorado, Boulder, Colorado 80309, USA\\
$^{2}$Departamento de F\'{\i}sica, Universidad de Extremadura, E-06071 Badajoz, Spain}

\begin{abstract}
A linear stability analysis of the Navier-Stokes (NS) granular hydrodynamic equations is performed to determine the critical length scale for the onset of vortices and clusters instabilities in granular dense binary mixtures. In contrast to previous attempts, our results (which are based on the solution to the inelastic Enskog equation to NS order) are not restricted to nearly elastic systems since they take into account the complete nonlinear dependence of the NS transport coefficients on the coefficients of restitution $\alpha_{ij}$. The theoretical predictions for the critical length scales are compared to molecular dynamics (MD) simulations in flows of strong dissipation ($\alpha_{ij}\geq 0.7$) and moderate solid volume fractions ($\phi\leq 0.2$). We find excellent agreement between MD and kinetic theory for the onset of velocity vortices, indicating the applicability of NS hydrodynamics to polydisperse flows even for strong inelasticity, finite density, and particle dissimilarity.
\end{abstract}

\draft 
\date{\today}
\maketitle

Although hydrodynamics is frequently used to describe rapid granular flows, there are still some open questions about the domain of validity of this description \cite {G03}. As for ordinary fluids \vicente{at moderate densities}, the constitutive equations for the irreversible fluxes and the forms of the transport coefficients can be derived from the revised Enskog kinetic theory (RET) \cite{BDS97} conveniently adapted to account for the dissipative dynamics. The derivation of such fluxes from the corresponding kinetic equation assumes the existence of a hydrodynamic regime where all space and time dependence of the distribution function only occurs through the hydrodynamic fields (\emph{normal} solution). The Chapmanû-Enskog expansion \cite{CC70} around the homogeneous cooling state (HCS) provides a constructive means to obtain an approximation to such a normal solution for states in which spatial gradients are not too large. A first-order Chapman-Enskog expansion provides the Navier-Stokes (NS) hydrodynamic equations and also explicit expressions for the corresponding transport coefficients, which are defined as functions of the coefficient of restitution and other system parameters. However, there are still some concerns regarding the transition from kinetic theory to hydrodynamics beyond the quasielastic limit \cite{G03}. The reason for this concern resides in the fact that the inverse of the cooling rate, which measures the rate of energy loss due to collisional dissipation, introduces a new timescale not present for elastic collisions. The variation of the granular temperature over this new timescale is faster than over the usual hydrodynamic timescale. However, as the inelasticity increases, it is possible that the system could lack a separation of time scales between the hydrodynamic and the pure kinetic excitations such that there is no \emph{aging} to hydrodynamics or, in the language of kinetic theory, there is no normal solution at finite dissipation.

Strictly speaking, to definitively address the validity of hydrodynamics for dissipative systems, the complete spectrum of the (linearized) Enskog-Boltzmann collision operator must be known. More specifically, knowing this spectrum allows one to see if the hydrodynamic modes (density, velocity, and temperature) decay more slowly than the remaining kinetic excitations at large times. To the best of our knowledge, this has been only accomplished \cite{DB03} by considering a simple model of the Boltzmann collision operator. On the other hand, the complex mathematical structure of the RET makes an exact solution to this kinetic equation intractable, even for studying the relaxation of small spatial perturbations of the HCS. An alternative approach is to compare the approximate theoretical solutions to the RET (obtained from the Chapman-Enskog method by assuming the validity of a normal solution) with numerical solutions to the RET obtained via the DSMC method \cite{B94}  or MD simulations \cite{AT89}. The latter does not utilize the RET and thus provides an even stronger test for the validity of the theory. The determination of the critical length scale $L_\text{c}$ for the onset of instabilities (which is generally driven by the transversal shear mode) in freely cooling flows offers one of the best opportunities to assess NS hydrodynamics. This kind of instability, which can be traced to the dissipative nature of collisions, is perhaps the most striking phenomenon that makes dissipative flows so distinct from ordinary (elastic) gases \cite{GZ93,BSSP04,BLN13,PJDR14}. Moreover, the comparison between kinetic theory and computer simulations for the critical size can be considered as a clean gauge of the former since both approaches (theory and simulation) are restricted to the linear regime where the deviations of the hydrodynamic fields from their values in the HCS are small.

The accuracy of the prediction of $L_\text{c}$ given by kinetic theory has been verified for a low-density \emph{monodisperse} granular gas by DSMC \cite{BRM98} and more recently by MD simulations for a \vicente{granular fluid at moderate density}  \cite{Peter11}. In both cases, the theoretical predictions for the critical size compare well with computer simulations even for strong dissipation. On the other hand, the results for polydisperse granular systems (namely, when the system is constituted by grains of different masses, sizes and coefficients of restitution) are more scarce. Polydispersity introduces phenomena that have no counterpart in monodisperse flow but may dramatically influence system behavior, such as species segregation \cite{segregation}. To the best of our knowledge, the only comparison for shearing instability for binary systems has been recently carried out in Ref.\ \cite{BR13} in the \emph{dilute} limit case where the collisional contributions (due to density effects) to the NS transport coefficients are neglected. As for Ref.\ \cite{BRM98}, theory compares well with the DSMC simulations of the Boltzmann equation. However, given that most of solids flows present in nature are dense and polydisperse, a proper theoretical framework for these systems is critical to obtain an accurate description of practical particle flows.

The aim of this Rapid Communication is to assess the ability of hydrodynamics to predict $L_\text{c}$ via a comparison with MD simulations in a binary granular mixture at moderate density. The theoretical results are based on a recent solution to the RET \cite{GDH07} that takes into account the nonlinear dependence of the transport coefficients on dissipation. This Chapman-Enskog solution \cite{GDH07} differs from some previous theoretical attempts for dense granular flows \cite{dense} that were obtained for quasielastic particles and so, none of the transport coefficients depend on the coefficients of restitution. Thus, the present theory subsumes all previous analysis \cite{BRM98,Peter11,BR13,dense}, which are recovered in the appropriate limits. Given that MD simulations avoid any assumptions inherent in the kinetic theory (e.g., molecular chaos) or approximations made in solving the RET by means of the DSMC method, the comparison between kinetic theory and MD simulations carried out here can be considered as the most stringent quantitative assessment of kinetic theory to date for conditions of practical interest. In this context, the results reported in this Rapid Communication provides strong evidence of the reliability of hydrodynamics for a wide range of densities and inelasticities in a quite complex (polydisperse) system.

We consider a binary mixture of \emph{inelastic}, smooth, hard spheres of masses $m_{1}$ and $m_{2}$, and diameters $\sigma_{1}$ and $\sigma_{2}$. The inelasticity of collisions among all pairs is characterized by three
independent constant coefficients of normal restitution $\alpha _{ij}$. At a kinetic level, the relevant information on the state of the mixture is given through the one-particle distribution functions which obey the RET. From it, one can derive the NS hydrodynamic equations for the \emph{granular} binary mixture with explicit expressions for the hydrostatic pressure, the cooling rate and the complete set of transport coefficients. The detailed form of the above quantities can be found in Ref.\ \cite{GDH07}. As for ordinary fluids, all these quantities (which have been approximately obtained by considering the leading terms in a Sonine polynomial expansion) are given in terms of the mole fraction $x_1=n_1/(n_1+n_2)$ ($n_i$ being the number density of species $i$), the mass ratio $m_1/m_2$, the size ratio $\sigma_1/\sigma_2$, the solid volume fraction $\phi$, and the coefficients of restitution $\alpha_{ij}$.

The hydrodynamic equations admit a nontrivial solution which corresponds to the so-called HCS. It describes a uniform state ($\nabla x_{1\text{H}}=\nabla n_{\text{H}}=\nabla T_{\text{H}}=0$) with vanishing flow field ($\mathbf{U}=\mathbf{0}$) and a granular temperature $T_\text{H}$ decreasing monotonically in time, namely, $
\left( \partial_{t}+\zeta_\text{H}^{(0)} \right) T_\text{H}=0$. Here, the subscript $\text{H}$ denotes the homogeneous state, $n_{\text{H}}=n_{1\text{H}}+n_{2\text{H}}$, and $\zeta_\text{H}^{(0)}\propto \sqrt{T_\text{H}}$ is the zeroth-order contribution to the cooling rate. However, it is well known \cite{GZ93} that the HCS is unstable with respect to long enough wavelength perturbations. To obtain quantitative estimates on the first stages of this instability, a (linear) stability analysis of the NS hydrodynamic equations with respect to the HCS can be performed. As usual, we assume that the deviations
$\delta y_{\beta}(\mathbf{r},t)=y_{\beta}(\mathbf{r},t)-y_{\text{H}\beta
}(t)$ are small, where $\delta y_{\beta}(\mathbf{r},t)$ denotes the deviation of $\{x_{1}, n, \mathbf{U}, T\}$ from their values
in the HCS. The resulting equations are then written in dimensionless form by using the (dimensionless) time $\tau=\frac{1}{2}\int_{0}^{t'}\nu_\text{H}(t')dt'$ and the (dimensionless) length $\boldsymbol{\ell}= \frac{1}{2}\frac{\nu_H(t)}{v_H(t)}\mathbf{r}$. Here, $v_{\text{H}}(t)=\sqrt{T_{\text{H}}(t)/\overline{m}}$, $\overline{m}=(m_1+m_2)/2$ and  $\nu_\text{H}=(8\pi^{(d-1)/2}/(d+2)\Gamma(d/2))n_\text{H}\sigma_{12}^{d-1}v_\text{H}$, where $\sigma_{12}=(\sigma_1+\sigma_2)/2$.

A set of Fourier transformed dimensionless variables are introduced by $\rho_{1,\mathbf{k}}=\delta x_{1\mathbf{k}}/x_{1\text{H}}, \rho_{\mathbf{k}}=\delta n_{\mathbf{k}}/n_{\text{H}}, \mathbf{w}_{\mathbf{k}}=\delta \mathbf{u}_{\mathbf{k}}/v_{\text{H}}$, and $\theta_{\mathbf{k}}=\delta T_{\mathbf{k}}/T_{\text{H}}$, where $\delta y_{\mathbf{k}\beta}\equiv \left\{\rho_{1,\mathbf{k}}, \rho_{\mathbf{k}}, \mathbf{w}_{\mathbf{k}}, \theta_{\mathbf{k}}\right\}$ is defined as
\begin{equation}
\delta y_{\mathbf{k}\beta}(\tau)=\int d\boldsymbol{\ell}\;e^{-i\mathbf{k} \cdot \boldsymbol{\ell}}
\delta y_{\beta}(\boldsymbol{\ell},\tau).
\label{5}
\end{equation}
As MD simulations carried out in this work clearly show, the origin of the instability is associated with the transversal components of the velocity field ${\bf w}_{{\bf k}\perp}={\bf w}_{{\bf k}}-({\bf w}_{{\bf k}}\cdot
\widehat{{\bf k}})\widehat{{\bf k}}$. As such, $L_\text{c}\equiv L_\text{vortex}$, where $L_\text{vortex}$ is the critical length scale for velocity vortex instabilities. As expected \cite{BRM98,Peter11,BR13}, the $d-1$ shear (transversal) modes ${\bf w}_{{\bf k}\perp}$ decouple from the other four longitudinal modes, greatly simplifying the theoretical analysis of the onset of instability. The evolution equation of ${\bf w}_{{\bf k}\perp}$ is
\begin{equation}
\label{6}
\frac{\partial {\bf w}_{{\bf k}\perp}}{\partial \tau}+\left(\frac{1}{2}\eta^*
k^2-\zeta_0^*\right){\bf w}_{{\bf k}\perp}=0,
\end{equation}
where $\eta^*\equiv (\nu_\text{H}\eta_\text{H})/(\rho_\text{H}v_\text{H}^2)$ is the (dimensionless) shear viscosity of the mixture and $\zeta_0^*\equiv \zeta_{\text{H}}^{(0)}/\nu_{\text{H}}$ \cite{GDH07}. Here, $\rho_\text{H}=m_1 n_{1\text{H}}+m_2 n_{2\text{H}}$ is the total mass density. The solution to Eq.\ \eqref{6} is ${\bf w}_{{\bf k}\perp}({\bf k}, \tau)={\bf w}_{{\bf k}\perp}(0)\exp[s_{\perp}(k)\tau]$, where $s_{\perp}(k)=\zeta_0^*-\frac{1}{2}\eta^* k^2$. This identifies a critical wave number $k_{\perp}^c=\sqrt{2\zeta_0^*/\eta^*}$ such that a linear excitation of the (scaled) transversal velocity with $k<k_{\perp}^c$ grows in time.
\begin{figure}
\includegraphics[width=0.9\columnwidth]{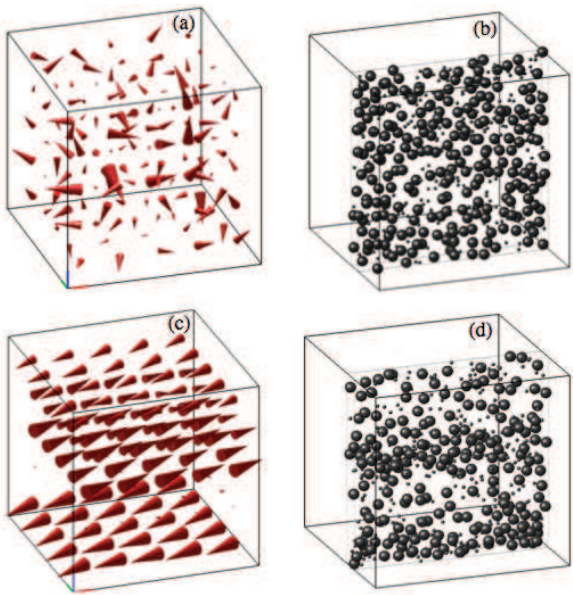}
\caption{(Color online) Visualizations from a MD simulation of an equimolar mixture ($x_1=0.5$) with $m_1/m_2=2$, $\sigma_1/\sigma_2=3$, $\phi =0.2$, and $\alpha=0.7$ of (a) stable, coarse-grained velocity field at 5 collisions per particle (or "cpp"), (b) stable particle positions at 5 cpp, (c) unstable, coarse-grained velocity field at 400 cpp, and (d) cluster systems at 400 cpp. A cell size of $L/5$ is used for local velocity averaging.}
\label{fig1}
\end{figure}

Since the simulations made here consider periodic boundary conditions, the smallest allowed wave number is $2\pi/L$, where $L$ is the system length. Hence, for given values of the parameters of the mixture (masses, diameters, composition, coefficients of restitution and volume fraction), we can identify a critical length $L_\text{vortex}$ such that the system becomes \emph{unstable} when $L>L_\text{vortex}$. The value of $L_\text{vortex}$ is
\begin{equation}
\label{7}
L_\text{vortex}=\frac{d+2}{2\sqrt{2}}\frac{\Gamma \left(\frac{d}{2}\right)}{\pi^{(d-3)/2}}\sqrt{\frac{\eta^*}{\zeta_0^*}}
\left(n_\text{H}\sigma_{12}^{d-1}\right)^{-1}.
\end{equation}

In order to assess the accuracy of the theoretical predictions, we have performed MD simulations. A cubic, periodic domain of length $L$ that consists of a total number of $N$ spheres ($d=3$) is simulated via hard-sphere MD \cite{AT89}. The parameter space over which Eq. \eqref{7} has been verified is the mole fraction $x_1$, the mass ratio $m_1/m_2$, the ratio of diameters $\sigma_1/\sigma_2$, the solid volume fraction $\phi=(\pi/6)(n_1\sigma_1^3+n_2\sigma_2^3)$ and the (common) coefficient of restitution $\alpha\equiv\alpha_{11}=\alpha_{22}=\alpha_{12}$. Two different values of the solid volume fraction $\phi$ have been considered here, $\phi=0.1$ and $\phi=0.2$, both representing a \vicente{granular fluid with moderate density}. In addition, three different values of $\alpha$ have been studied: $\alpha=0.9$ (weak dissipation), $\alpha=0.8$ (moderate dissipation) and $\alpha=0.7$ (strong dissipation). To determine the stability of a given simulation with respect to velocity vortices, we have used a Fourier analysis \cite{Peter13}. Specifically, we use an integrated Fourier transform of the momentum field to assess the magnitude of contributions to given wavelengths. A monotonic increase in the magnitude of contribution with respect to wavelength corresponds to a homogeneous flow. Relatively large contributions (or excitations) at small wavelengths ($2\pi/L$ or $4\pi/L$ in the HCS) correspond to velocity vortices, organized structures in the momentum field.

\begin{figure}
{\includegraphics[width=0.51\columnwidth]{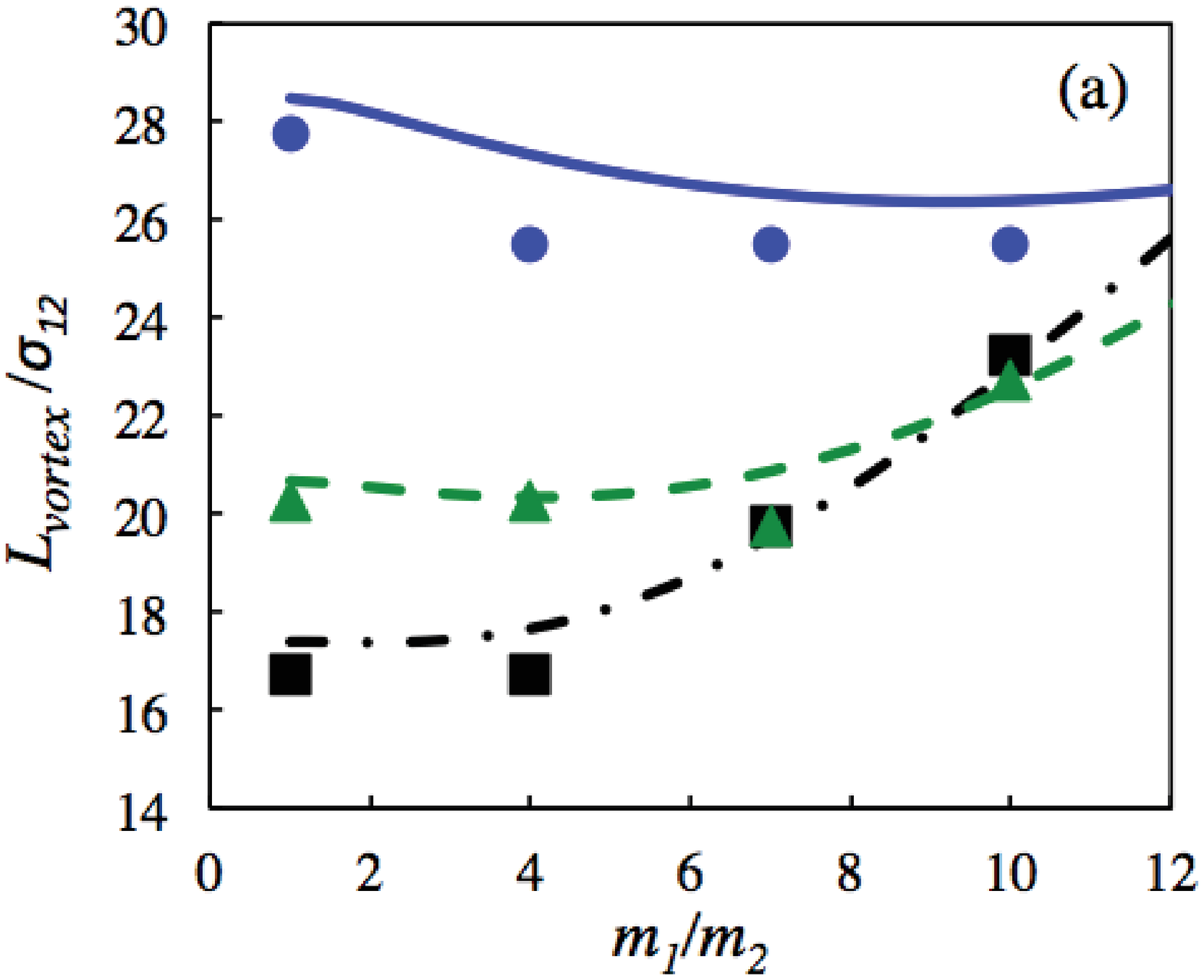}}
{\includegraphics[width=0.47\columnwidth]{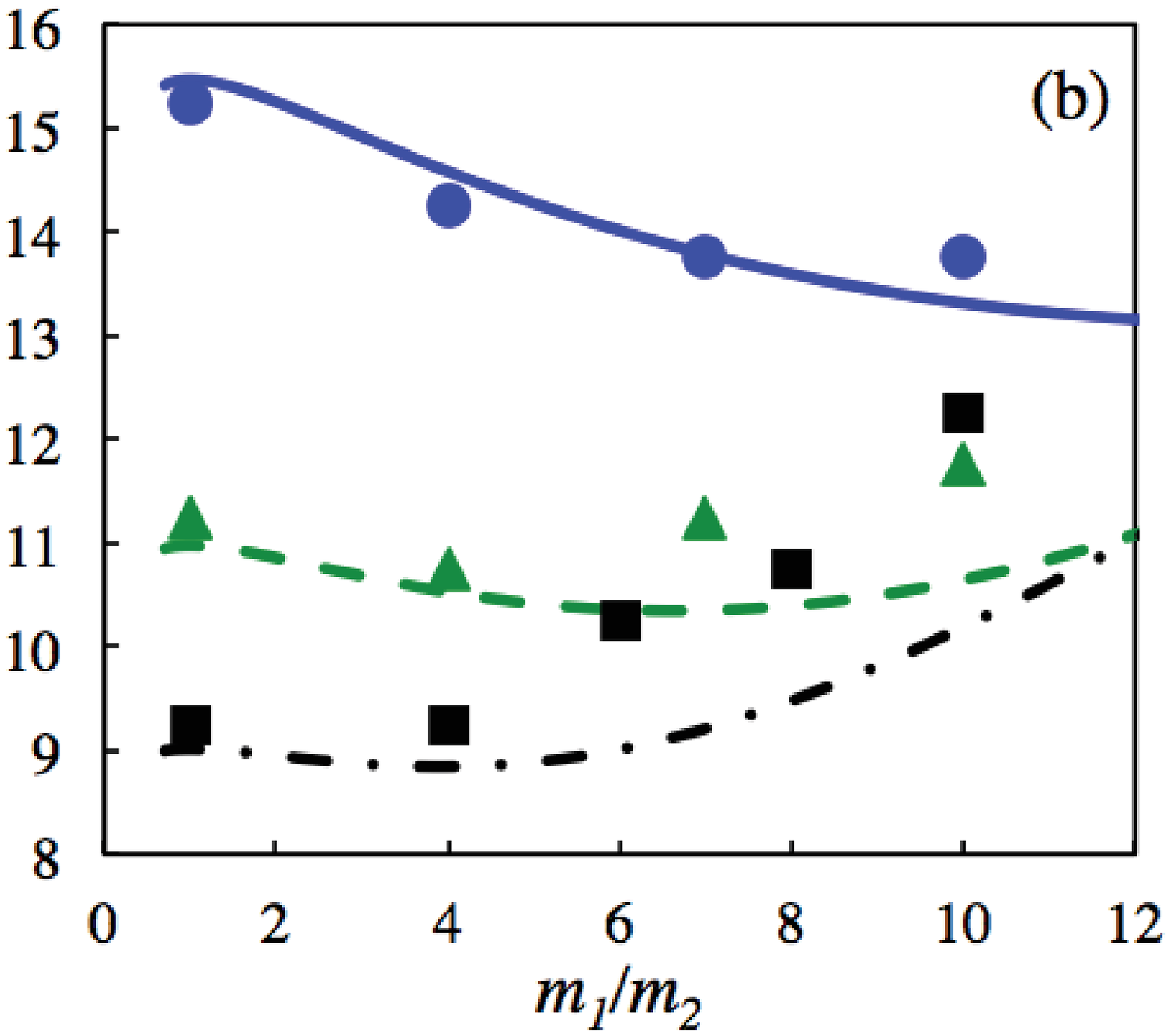}}
\caption{(Color online) Critical length scale for velocity vortices as a function of the mass ratio $m_1/m_2$ with $x_1=0.1$, $\sigma_1/\sigma_2=1$ for (a) $\phi=0.1$ and (b) $\phi=0.2$. The data points correspond to MD simulations while the lines are the theoretical predictions given by Eq.\ \eqref{7}. (Blue) circles/solid line, (red) triangles/dashed line, and (black) squares/dot-dashed line correspond to $\alpha=0.9$, $\alpha=0.8$, and $\alpha=0.7$, respectively. Error ranges are the size of the data points and are omitted.
\label{fig2}}
\end{figure}

For each set of particle parameters simulated, a critical dimensionless length scale exists that distinguishes stable systems from ones unstable to velocity vortices. To determine this critical scale, we have considered 24 replicate simulations (that only differ in initial conditions) for a range of domain length scales. If any one of the 24 replicates is unstable, the corresponding $L/\sigma_{12}$ is considered unstable. Thus, a range for $L_\text{vortex}/\sigma_{12}$ can be determined where the higher $L/\sigma_{12}$ (i.e., upper bound) of this range is unstable and the smaller $L/\sigma_{12}$ (i.e., lower bound) is stable. As an illustration, Fig.\ \ref{fig1} shows snapshots of velocity and concentration fields. Small systems will remain stable (see Figs. 1(a) and 1(b) for homogeneous velocity field and particle positions, respectively), while instabilities (see Figs. 1(c) and 1(d) for vortices and clusters, respectively) will manifest in large systems after long times.

Next, the (linear) hydrodynamic predictions of $L_\text{vortex}$ given by Eq.\ \eqref{7} are compared
to results from MD simulations. Figure \ref{fig2} shows $L_\text{vortex}/\sigma_{12}$ as a function of the mass ratio
and coefficient of restitution with $\sigma_1/\sigma_2=1$, and $x_1=0.1$. It is quite apparent that Fig.\ 2(a) ($\phi=0.1$) shows excellent agreement between hydrodynamics and MD simulations throughout the parameter space studied, even for significant
dissipation in combination with large mass ratios. For moderate densities ($\phi=0.2$) and dissipation ($\alpha=0.9$)
(see Fig.\ 2(b)), excellent agreement is observed up to quite extreme mass ratios ($m_1/m_2=10$), while for
higher dissipation ($\alpha=0.7$), strong agreement is observed to significant mass ratios ($m_1/m_2=4$).
An interesting qualitative agreement is also displayed in Fig. \ref{fig2}. For both MD simulations and hydrodynamics,
the $L_\text{vortex}/\sigma_{12}$ predictions for $\alpha=0.7$ and $\alpha=0.8$ begin to converge for large mass ratios
and eventually crossover. Figure 3(a) shows $L_\text{vortex}/\sigma_{12}$ as a function of the ratio of diameters with $x_1=0.5$, $m_1/m_2=2$, and $\phi=0.2$. Excellent agreement is observed throughout the conditions studied. Figure 3(b) shows the critical size as a function of species composition $x_1$ for a relatively large mass ratio ($m_1/m_2=6$) with $\sigma_1/\sigma_2=1$, and $\phi=0.2$. We observe that even in the extreme case of small composition ($x_1=0.1$) and large dissipation ($\alpha=0.7$), hydrodynamics and MD deviate by less than roughly 10\%. For moderate dissipation ($\alpha=0.9$), we observe that the theory compares very well with MD for all species compositions.

\begin{figure}
{\includegraphics[width=0.51\columnwidth]{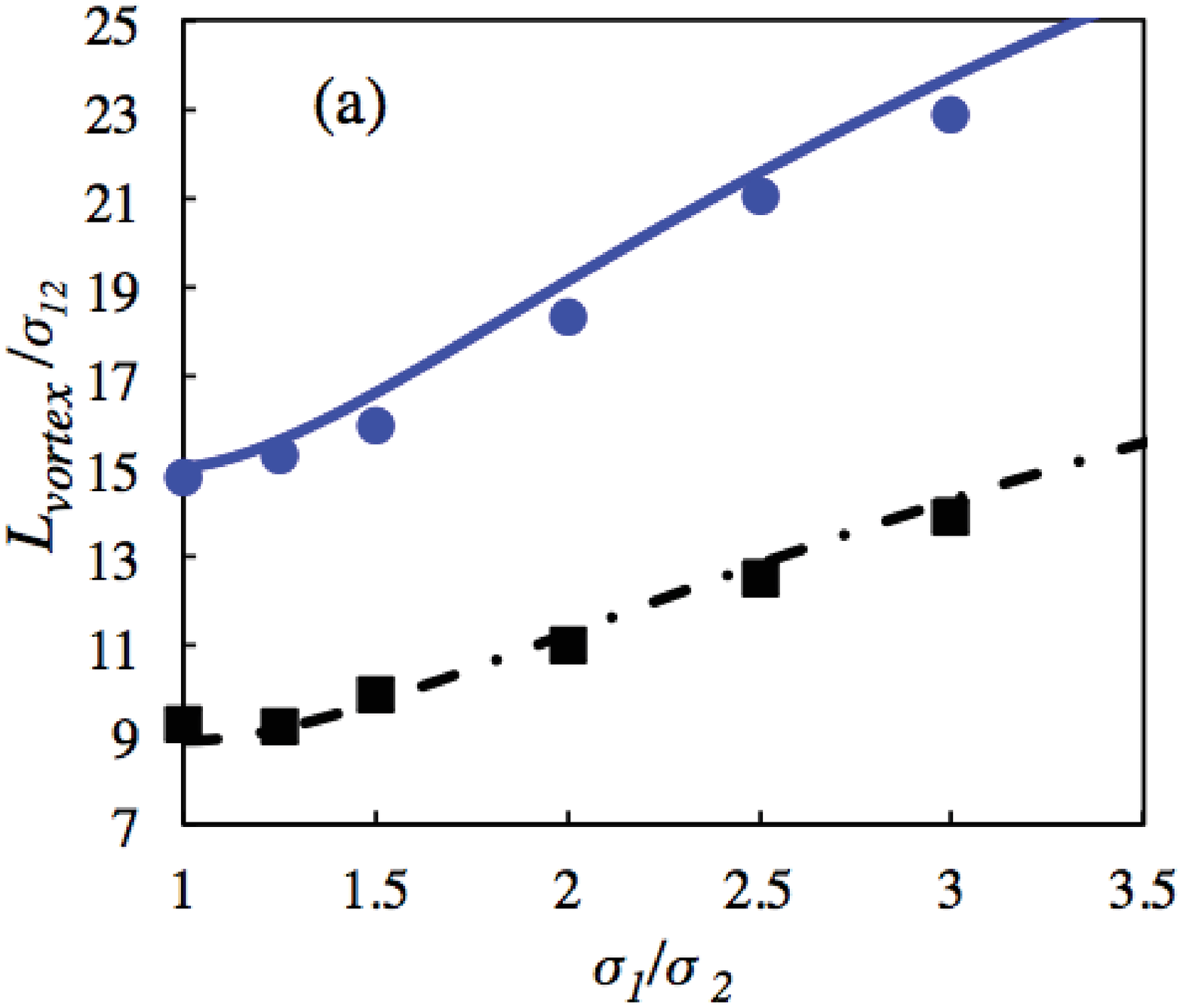}}
{\includegraphics[width=0.47\columnwidth]{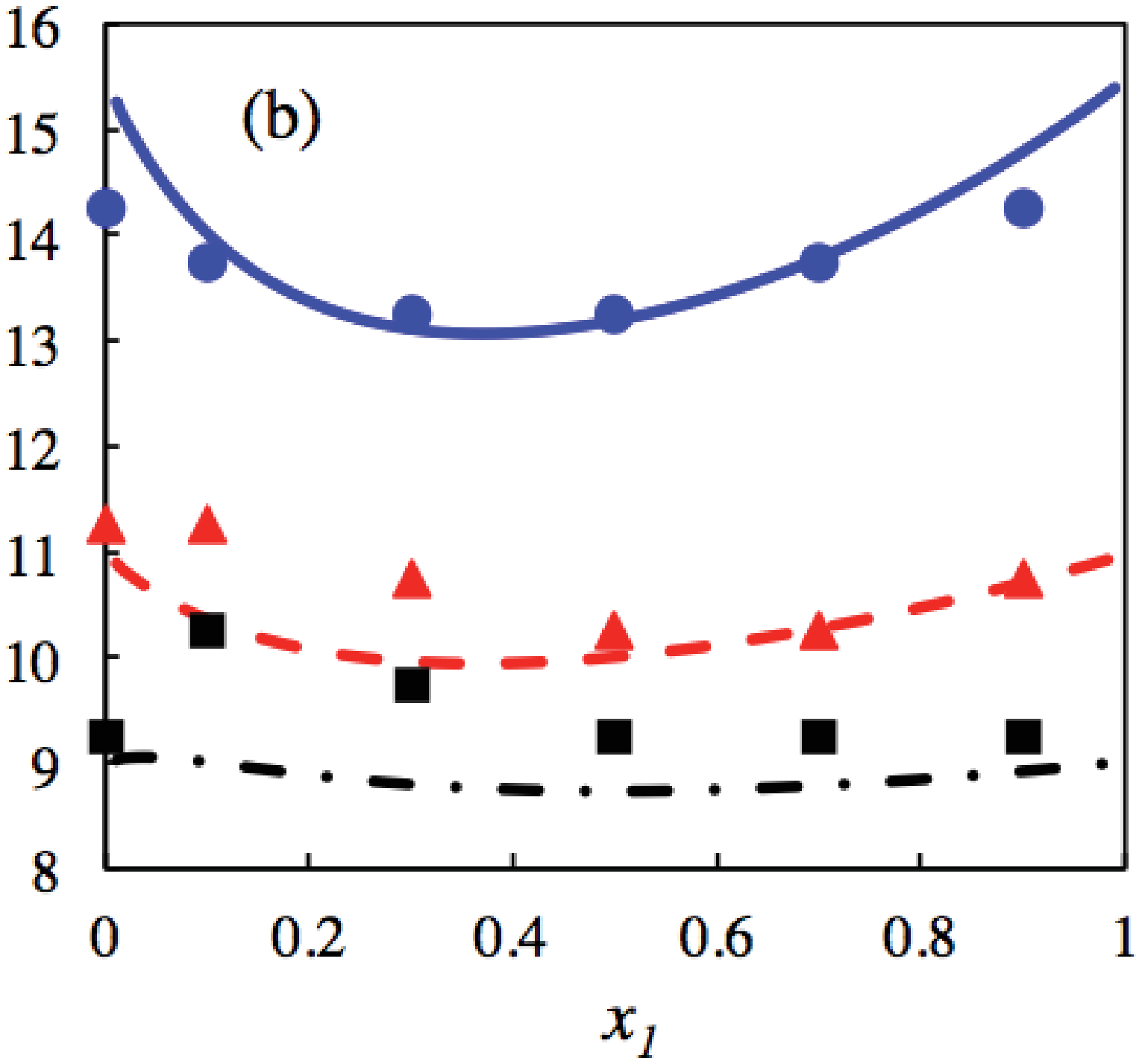}}
\caption{(Color online) Critical length scale for velocity vortices as a function of (a) the ratio of diameters $\sigma_1/\sigma_2$ with $m_1/m_2=2$,$x_1=0.5$, and  $\phi=0.2$ and (b) the mole fraction $x_1$ with $m_1/m_2=6$, $\sigma_1/\sigma_2=1$, and $\phi=0.2$. The meaning of  symbols and lines is the same as that of Fig.\ \ref{fig2}.
\label{fig3}}
\end{figure}

We see that the comparison carried out in Figs.\ \ref{fig2} and \ref{fig3} for $L_\text{vortex}$ shows in general an excellent agreement between theory and simulation when physical properties of particles are similar, while only good (at worst ~20\% error) for the most extreme conditions studied (e.g., Fig. 2(b); $m_1/m_2=10$, $x_1=0.1$, $\phi= 0.2$, and $\alpha=0.7$). On the other hand, based on the results derived from the RET for ordinary fluid mixtures \cite{KCL87}, one would expect that the accuracy of the first Sonine approximation to the shear viscosity $\eta^*$ (which is the transport coefficient involved in $L_\text{vortex}$, see Eq.\ \eqref{7}) would decrease as the mass ratio becomes more disparate. In this context, the leading order truncation of the Sonine polynomial expansion may be responsible for the discrepancies found between theory and MD rather than assumptions inherent to the RET, such as the absence of velocity correlations (molecular chaos hypothesis). In fact, recent results \cite{GV09} for the tracer diffusion coefficient $D$ have shown that in general the second Sonine approximation to $D$ improves significantly the prediction of the first Sonine solution, especially for high dissipation and/or extreme mass or diameter ratios. The inclusion of the second-order Sonine corrections to $\eta^*$ could mitigate part of the discrepancies observed here, especially in the case of quite extreme values of the mass ratio.

In summary, the comparison addressed here between the predictions of linear hydrodynamics (derived from the RET) and discrete-particle computer simulations provides the most stringent test to date for hydrodynamic description of multicomponent granular fluids. This hydrodynamic description continues to be a source of controversy, from the appropriateness of the RET (which is derived under the molecular chaos assumption) for flows \vicente{with moderate densities} to the appropriateness of the NS equations. Here, we use MD simulations (which do not rely on any of the above assumptions) as our ideal data set. The system we examine is complex; hydrodynamic instabilities in a transient, polydisperse system \vicente{at moderate density} with significant dissipation levels. The good agreement found in this paper between the predictions of linear hydrodynamics (with the NS transport coefficients derived form the RET) and MD simulations must be considered as a nontrivial example of the reliability of hydrodynamics as a quantitative predictive tool for moderate dense and highly dissipative granular binary mixtures. Therefore, although the theoretical method used here (Chapman-Enskog) is formal and does not strictly establishes the existence of hydrodynamics, our results (theory and MD simulations) clearly indicate that the granular temperature can be still considered as a \emph{slow} hydrodynamic variable such that the \emph{normal} solution to the RET is still applicable. It is also worth noting that recent work \cite{MZBGDH13} has shown that NS order equations for monodisperse granular systems perform well even in the presence of quite large spatial gradients (whereas small gradients are present at the onset of the vortex instability examined here). This finding echoes what is also true for rarefied gases, i.e., the NS approximation has a much wider range of applicability than would be expected from a strict interpretation of its assumption. Accordingly, the current work on polydispersity and recent work \cite{MZBGDH13} on higher-order effects implies that the NS hydrodynamic description is much more robust than previously considered possible.

Hydrodynamic descriptions derived from kinetic theory are critical tools in the description of numerous industrial processes involving solid particles. These descriptions are now standard features of commercial, multiphase computational fluid dynamics codes (like Fluent) and open-source, research codes (like MFIX). Given that such codes rely on accurate expressions for the transport coefficients, it is evident that the results displayed in the present paper are of great valuable not only for the granular physics community working on kinetic theory but also for more applied scientists interested in more practical problems (e.g., biomass gasification, mixing of pharmaceutical powders, heat exchange in concentration solar power plants, synthesis of fine chemicals, pollution control, and ejection of lunar soil from rocket landings). Because of this pervasiveness, the applicability of hydrodynamics to complex polydisperse systems should resonate throughout the fluid dynamics community.

P.P.M. and C.M.H. would like to acknowledge the funding provided by the American Chemical Society (Grant PRF 50885-ND9) and the National Science Foundation (Grant CBET-1236157). The research of V.G.has been supported by the Spanish Government through Grant Nos. FIS2010-12587, partially financed by FEDER funds and by Junta de Extremadura (Spain) through Grant No.\ GRU10158.

\end{document}